\documentclass[submission,copyright,creativecommons]{eptcs}

\usepackage[utf8]{inputenc}
\usepackage{amsmath}
\usepackage{enumerate}
\usepackage{mathpartir}
\usepackage{mathtools}
\usepackage{pmboxdraw}
\usepackage[T1]{fontenc}

\usepackage[normalem]{ulem}
\usepackage{mathpartir}
\usepackage{proof}
\usepackage{mathtools}
\usepackage{longtable}
\usepackage{supertabular}
\usepackage{ stmaryrd }

\usepackage{comment}
\usepackage{color}   
\hypersetup{
    colorlinks,
    citecolor=black,
    filecolor=black,
    linkcolor={blue!50!black},
    urlcolor=black
}

\usepackage{tikz} 

\usepackage{diagbox}
\usepackage{pdflscape}

\usepackage{makecell}
\usepackage{changepage}
\usepackage{xcolor}
\usepackage{multicol}
\usepackage{multirow}
\usepackage{listings}
\usepackage{etoolbox}
\usepackage{environ}
\usepackage{newunicodechar}
\usepackage{bm}
\usepackage{vwcol}  

\usepackage{framed}
\usepackage{cprotect}

\newunicodechar{∞}{\ensuremath{\mathnormal\infty}}
\newunicodechar{♯}{\ensuremath{\mathnormal\sharp}}
\newunicodechar{⌊}{\ensuremath{\mathnormal{\lfloor}}}
\newunicodechar{⌋}{\ensuremath{\mathnormal{\rfloor}}}
\newunicodechar{≟}{\ensuremath{\mathnormal{\stackrel{?}{=}}}}
\newunicodechar{≡}{\ensuremath{\mathnormal{\equiv}}}

\newcommand{\mt}[1]{\mathop{\operatorname{\mathtt{{#1}}}}}

\newcommand\hra\hookrightarrow

\newcounter{numbered}
\newenvironment{numbered}{
    \setcounter{numbered}{0}
    }{
    \setcounter{numbered}{0}
    }

\newcommand{\previousNumber}{\the\numexpr\value{numbered}-1\relax }

\usepackage[numbers]{natbib}

\newtoggle{journalversion}
\toggletrue{journalversion}

\iftoggle{journalversion}{
}{
}

\iftoggle{journalversion}{
    \includeonly{}
}{
    \includeonly{appendix}
}

\newcommand{\colf}{CoLF${}^\omega$}
\newcommand{\colfone}{CoLF${}^\omega_1$}
\title{CoLF Logic Programming as Infinitary Proof Exploration}
\author{Zhibo Chen
\institute{Carnegie Mellon University}
\email{zhiboc@andrew.cmu.edu}
\and
Frank Pfenning
\institute{Carnegie Mellon University}
\email{fp@cs.cmu.edu}
}

\def\titlerunning{\colfone Logic Programming}
\def\authorrunning{Z. Chen, F. Pfenning}

\begin{document}

\maketitle

\begin{abstract}
  Logical Frameworks such as Automath \citep{DeBruijn68} or 
  LF \citep{Harper93jacm} were originally conceived as
  metalanguages for the specification of foundationally uncommitted deductive
  systems, yielding generic proof checkers.  Their high level of abstraction was
  soon exploited to also express algorithms over deductive systems such as
  theorem provers, type-checkers, evaluators, compilers, proof transformers,
  etc. in the paradigm of computation-as-proof-construction.  This has been
  realized in languages such as $\lambda$Prolog \citep{Miller91apal} or Elf \citep{Pfenning91book} based on backward
  chaining, and LolliMon \citep{Lopez05ppdp} or Celf \citep{SchackNielsen08ijcar},
   which integrated forward chaining.

  None of these early frameworks supported the direct expression of infinitary
  objects or proofs, which are available in the recently developed \colf \citep{Chen23arxivInfLF}.  In
  this work-in-progress report, we sketch an approach to
  computation-as-proof-construction over the first-order fragment of \colf\ 
  (called \colfone) that already includes infinitary objects and proofs.  A key
  idea is the interpretation of logic variables as communication channels and
  computation as concurrent message-passing.  This is realized in a concrete
  compiler from \colfone\ to Sax, a proof-theoretically inspired parallel
  programming language based on the proof-reduction in the semi-axiomatic
  sequent calculus \citep{DeYoung20fscd}.
\end{abstract}







 


 








\section{Introduction}

Consider the set of natural numbers inductively generated by 
symbols \verb$s$ and \verb$z$. The following two rules define the addition
operation on natural numbers.
\begin{mathpar}
\inferrule{
}{
  \mt{add}\, \mt{z}\, \mt{A}\, \mt{A}
}\mt{add\_z}

\inferrule{
  \mt{add}\, \mt{A}\, \mt{B}\, \mt{C} \\
}{
  \mt{add}\, (\mt{s}\, \mt{A})\,\mt{B}\, (\mt{s}\, \mt{C})
}\mt{add\_s}
\end{mathpar}

The adequacy of representation dictates that a derivation of \verb$add A B C$
exists if and only if $A + B = C$.  We can represent both the natural numbers
and the addition relation in LF as follows.

\begin{verbatim}
nat: type.
z : nat.
s : nat -> nat.

add: nat -> nat -> nat -> type.  %mode add + + -.
add_z : add z A A. 
add_s : add A B C -> add (s A) B (s C).
\end{verbatim}

The computational interpretation of this signature in Twelf proceeds by
searching for a derivation of of \verb$add A B C$, given \verb$A$ and \verb$B$.  Mode
checking guarantees that if the first two arguments are ground and proof search
succeeds, then the third argument will also be ground.  In this example, there
is a unique \verb$C$ such that the relation \verb$add A B C$ holds, but in general
proof search may backtrack and thereby either fail to terminate or enumerate
multiple solutions.

Backward-chaining proof search in this style presents multiple difficulties in
the infinitary settings of \colf.  One is termination if the inputs are
infinitary.  Infinitary objects also interact poorly with backtracking because
we may never definitively fail.  Related is a problem with unification, which is
guaranteed to terminate only over rational (that is, circular) terms, while in
many applications either objects or proofs are not rational in this sense.

What we would like is a dynamics where outputs are computed incrementally from
inputs, evoking the image of transducers between streams (even if terms
generally have the shape of potentially infinite trees, not just streams).

In order to avoid backtracking, we use mode and uniqueness checking to
statically enforce that there will be at most one proof for each unique input.
As a (perhaps surprising) consequence we can then exploit and-parallelism
between multiple premises, reducing synchronization between them to
communication between shared variables.

For example, consider the following program that computes the product of two
conatural numbers (that is, potentially infinite numbers).  We continue
to use \verb$s$ and \verb$z$ as constructors.  Note that both \verb$add$
and \verb$mult$ are cotypes, because we want to allow infinitary derivations
for addition and multiplication.

\begin{verbatim}
conat: cotype.
z : conat.
s : conat -> conat.

add: conat -> conat -> conat -> cotype.  %mode add + + -.
add_z : add z A A.
add_s : add A B C -> add (s A) B (s C).

mult : conat -> conat -> conat -> cotype.  %mode mult + + -.
mult_z : mult z A z.
mult_s : mult A B C -> add B C D -> mult (s A) B D.
\end{verbatim}

Here, the two premises \verb$mult A B C$ and \verb$add B C D$ can be evaluated
in parallel, with some interesting flow of information.  For example, in a lazy
setting, where the output \verb$D$ is revealed step by step, we don't need to
evaluate \verb$mult A B C$ for the first \verb$B$ steps since \verb$C$ remains
unchanged.

In summary, we explore a logic programming language based on \colfone\ where
proofs are represented as infinitary terms, proof construction does not
backtrack, and premises of rules are evaluated in parallel using shared logic
variables for communication. 
In the rest of this paper, we give definitions of streams and stream transducers, 
and explain their operational semantics. 


\section{Simple Examples}

We give an overview of the intended meaning of CoLF logic programs 
through examples. The examples we primarily consider are streams of 
conatural numbers. The current semantics of CoLF handles only the 
coinductive fragment, contrary to the mixed inductive and coinductive case
\citep{Chen23arxivInfLF,Chen21ms}.

The data signature is given below, where \verb$s$ and \verb$z$ are the constructors 
of conatural numbers, and \verb$cons$ is the only constructor for streams.

\begin{verbatim}
conat: cotype.
z: conat.
s: conat -> conat.

stream: cotype.
cons: conat -> stream -> stream.
\end{verbatim}

We may use notational definitions to define several streams. Notational 
definitions are elaborated into relations during compilation. 
We define \verb$repeat n$ to be the stream that repeats the number $n$ infinitely, 
and \verb$up n$ to be the stream that counts up from $n$.
Notice that the syntax for the above definitions follows that of Twelf \citep{Pfenning98pstt}, where 
$\lambda$-abstractions are written using square brackets.
\begin{verbatim}
repeat : conat -> stream = [N] cons N (repeat N).
up : conat -> stream = [N] cons N (up (s N)).
\end{verbatim}

The definition of \verb$repeat$ is that \verb$repeat n$ is \verb$cons n (repeat n)$, 
a stream whose head is \verb$n$ and whose tail is the same stream.
The definition of \verb$up$ is that \verb$up n$ is \verb$cons n (up (s n))$, 
a stream whose head is \verb$n$ and whose tail is the stream that counts up from \verb$s n$.
The above two definitions are equivalent to the following relational definitions. 
The compiler performs this translation automatically as part of the compilation process.
We leave the universal quantification of the free variables (in uppercase)
implicit.


\begin{verbatim}
repeat : conat -> stream -> cotype.  %mode repeat + -.
repeat_def : repeat N R -> repeat N (cons N R).

up : conat -> stream -> cotype.  %mode up + -.
up_def : up (s N) U -> up N (cons N U).
\end{verbatim}

We can also view the relation as 
the following rules for constructing infinitary proofs.
\begin{mathpar}
  \inferrule{
    \mt{repeat}\, \mt{N}\, \mt{R} \\
  }{
    \mt{repeat}\, \mt{N}\, (\mt{cons}\, \mt{N}\, \mt{R})
  }(\mt{repeat\_def})

  \inferrule{
    \mt{up}\, (\mt{s}\,\mt{N})\, \mt{U} \\
  }{
    \mt{up}\, \mt{N}\, (\mt{cons}\, \mt{N}\, \mt{U})
  }(\mt{up\_def})
\end{mathpar}

The rules mimic the recursive definitions: an infinitary proof expansion of 
\verb$repeat N R$ will equate \verb$R$ with \verb$cons N (cons N (cons N ...))$,
and an infinitary proof expansion of \verb$up N U$ will equate \verb$U$ 
with \verb$cons N (cons (s N) (cons (s (s N)) ...))$. We show a partial expansion below.
\begin{mathpar}
  \inferrule{
  \inferrule{
  \inferrule{
    \mt{repeat}\, \mt{N}\, \mt{\dots} \\
  }{
    \mt{repeat}\, \mt{N}\, (\mt{cons}\, \mt{N}\, \mt{\dots})
  }
  }{
    \mt{repeat}\, \mt{N}\, (\mt{cons}\, \mt{N}\, (\mt{cons}\, \mt{N}\, \mt{\dots}))
  }
  }{
    \mt{repeat}\, \mt{N}\, (\mt{cons}\, \mt{N}\, (\mt{cons}\, \mt{N}\, (\mt{cons}\, \mt{N}\, \mt{\dots})))
  }

  \inferrule{
  \inferrule{
  \inferrule{
    \mt{up}\,(\mt{s}\, (\mt{s}\, (\mt{s}\, \mt{N})))\, \dots
  }{
    \mt{up}\, (\mt{s}\, (\mt{s}\, \mt{N}))\, (\mt{cons}\, (\mt{s}\, (\mt{s}\, \mt{N}))\, \dots)
  }
  }{
    \mt{up}\, (\mt{s}\, \mt{N})\, (\mt{cons}\, (\mt{s}\, \mt{N})\, (\mt{cons}\, (\mt{s}\, (\mt{s}\, \mt{N}))\, \dots))
  }
  }{
    \mt{up}\, \mt{N}\, (\mt{cons}\, \mt{N}\, (\mt{cons}\, (\mt{s}\, \mt{N})\, (\mt{cons}\, (\mt{s}\, (\mt{s}\, \mt{N}))\, \dots)))
  }
\end{mathpar}

In fact, we are able to run this program in our implementation by providing a main function. 
For example, we may be interested in seeing the evaluating \verb$up z$ and we may write 
a main definition (or relation) as follows.

\begin{verbatim}
main : stream = up z.
\end{verbatim}

The main function may also be defined as a relation. This process is automatically 
carried out by the compiler.
\begin{verbatim}
main : stream -> cotype.  %mode main -.
main_def : up z U -> main U.
\end{verbatim}

The compiler will then compile and perform infinitary proof expansion. It will stop after
the result is ground up to a predetermined depth. Here is the result from the compiler,
where \verb$...$ indicates a term that has not been computed yet.
\begin{verbatim}
(cons z 
(cons (s z) 
(cons (s (s z)) 
(cons (s (s (s z))) 
(cons (s (s (s (s z)))) ...)))))
\end{verbatim}

For space reasons, we truncated the output above. 
The effect of computing up to a certain term depth
 is more prominent when we compute the infinite stream of natural number $\omega$ as follows.

\begin{verbatim}
omega : conat = s omega.
main : stream = repeat omega.
\end{verbatim}

Notice here that the argument to \verb$repeat$ is also defined recursively, 
and the compiler correctly elaborates them into definitions.
\begin{verbatim}
omega : conat -> cotype.  %mode omega -.
omega_def : omega O -> omega (s O).
main : stream -> cotype.  %mode main -.
main_def : omega O -> repeat O R -> main R.
\end{verbatim}

Here, something interesting happens in the definition of \verb$main$. Since
\verb$omega O$ and \verb$repeat O R$ are two separate premises, they may be
evaluated in parallel. Moreover, although semantically \verb$omega O$ is
outputting a stream \verb$O$ to be read by \verb$repeat O R$, since
\verb$repeat$ is a recursive definition that does not look at the structure of
its first argument, there is no strict evaluation order on whether \verb$repeat$
or \verb$omega$ should be evaluated first.

Running the interpreter, we get the following stream as the result.
In this case, the stream is computed up to depth $5$, and we can see that 
the first element is computed up to depth $4$, the second element is computed up to depth $3$, and etc.
\begin{verbatim}
(cons (s (s (s (s ...)))) 
(cons (s (s (s ...))) 
(cons (s (s ...)) 
(cons (s ...) 
(cons ...)))))
\end{verbatim}

\section{More Complex Relations}

We now turn to more complex relations that include case analysis on 
input arguments to a relation.

As in the LF logical framework, term abstractions cannot perform a case analysis
on their arguments but may only use them ``parametrically''. To analyze the
input arguments, we need to define a relation and provide different branches for
each input argument case.

For example, consider the following addition relation defined on conatural 
numbers. The rules should be straightforward as we have seen these for several times
so far.

\begin{verbatim}
add: conat -> conat -> conat -> cotype.  %mode add + + -.
add_z : add z A A.
add_s : add A B C -> add (s A) B (s C).
\end{verbatim}

The core question is: what does this program mean computationally?

Rather than giving a full formal semantics, we explain informally the program
behavior of \verb$add X Y Z$.  The idea is to treat each argument as a channel
of communication. The mode declaration specifies whether channels are inputs
(mode $+$) or outputs (mode $-$).  For instance, the program \verb$add X Y Z$
operates on two input channels \verb$X$ and \verb$Y$, and one output channel
\verb$Z$.  The behavior of the program is specified by the clauses, directly
modeling infinitary proof construction.

The first clause \verb$add_z$ states that the program \verb$add X Y Z$ \emph{reads} 
the first input channel \verb$X$, and if the value read is \verb$z$, 
the program \emph{forwards} the input channel \verb$Y$ to output channel \verb$Z$.

The second clause \verb$add_s$ states that the program \verb$add X Y Z$ \emph{reads}
the first input channel \verb$X$. If the value read is \verb$s A$, for some input channel \verb$A$,
the program \emph{allocates} a fresh channel \verb$C$, and \emph{writes} to the output channel \verb$Z$ 
the value \verb$s C$, then we continue as the program \verb$add A Y C$.

The proposed informal semantics has the following properties:
\begin{enumerate}
  \item The program operates on list of input channels and a single output channel
  \item At each step, the program consists of interacting processes that may
    perform one of the following actions:
  \begin{enumerate}
    \item Read a value from an input channel
    \item Write a value to an output channel
    \item Forward an input channel to an output channel
    \item (During writing) Allocate fresh channels
    \item Spawn a new process (see stream processor example below)
    \item Continue as some process
  \end{enumerate}
  \item No backtracking (see below)
\end{enumerate}

The property of no backtracking states that at each program point, 
the process has at most one possible action. In other words, 
the clauses for a relation have to agree on the action taken at each step. 
In the case of \verb$add$ above, we see that the first action in both clauses 
should be to read from the first input channel \verb$X$. Then, based on different values of 
\verb$X$, each clause may take different actions. 
The compiler checks uniqueness during compilation, and signals an error if
more than one action is possible at some program point.

We now define the pointwise addition of two streams.
\begin{verbatim}
add_stream: stream -> stream -> stream -> cotype.  %mode add_stream + + -.
add_stream_def : add A B C -> add_stream R S T ->
  add_stream (cons A R) (cons B S) (cons C T).
\end{verbatim}

The clause \verb$add_stream_def$ states that 
the program \verb$add_stream X Y Z$ will carry out the following actions in order:
\begin{enumerate}
  \item Read from the first input channel \verb$X$ a stream \verb$cons A R$.
  \item Read from the second input channel \verb$Y$ a stream \verb$cons B S$.
  \item Allocate a fresh channel \verb$C$, and spawn a new process \verb$add A B C$.
  \item Allocate a fresh channel \verb$T$, and spawn a new process \verb$add_stream R S T$.
  \item Write to the output channel \verb$Z$ the value \verb$cons C T$.
\end{enumerate}

There are two things to notice here. First, we dictate that the program read the
first input channel \verb$X$ before reading the second input channel \verb$Y$.
Strictly speaking, this is not necessary in this case, but when we get to the
case of multiple clauses, this is needed because we would like to ensure unique
decision at every program point. Second, steps (3) (4) and (5) can happen
simultaneously, given that fresh channels \verb$C$ and \verb$T$ are allocated.
However, due to the fact that we do not allow backtracking, we still ensure a
sequential order of actions where the spawned processes can execute in parallel
with the main process.

With the help of stream addition, as an example, we may define the 
stream of even numbers by adding together two streams 
that count up from $0$.  Note here that \verb$even$ cannot be 
a constant definition but has to be a relation, because \verb$stream_add$
needs to analyze its arguments.

\begin{verbatim}
even : stream -> cotype.  %mode even -.
even_def : add_stream (up z) (up z) E -> even E.
\end{verbatim}

We may evaluate the \verb$even$ predicate through a \verb$main$ relation.
Since \verb$even$ is a relation, \verb$main$ also needs to be a relation, not 
a constant definition.

\begin{verbatim}
main : stream -> cotype.  %mode main -.
main_def : even E -> main E.
\end{verbatim}

And we get the expected result when executing the program.

\begin{verbatim}
(cons z 
(cons (s (s z)) 
(cons (s (s (s (s z)))) 
(cons (s (s (s (s (s (s z)))))) 
(cons (s (s (s (s (s (s ...))))))  ...)))))
\end{verbatim}

In a similar way, we may define the Fibbonacci stream by beginning with 
$0$ and $1$, and then adding the stream with its tail.

\begin{verbatim}
tail : stream -> stream -> cotype.  %mode tail + -.
tail_def : tail (cons N F) F.

fib : stream -> cotype.  %mode fib -.
fib_def : fib F -> tail F G ->
    add_stream F G H -> fib (cons z (cons (s z) H)).

main : stream -> cotype.  %mode main -.
main_def : fib F -> main F.
\end{verbatim}

Here, the channel represented by variable $F$ is an output
if \verb$fib$ and an input to two other processes (\verb$tail$
and \verb$add_stream$). 

We can test this program by running the interpreter.
\begin{verbatim}
(cons z 
(cons (s z) 
(cons (s z) 
(cons (s (s z)) 
(cons (s (s (s z))) 
(cons (s (s (s (s (s z))))) 
(cons (s (s (s (s (s (s (s (s z)))))))) ...)))))))
\end{verbatim}

Another interesting operation we can define is the integration operation (inspired by Budiu et al.~\citep{Budiu24sigmod}), which calculates 
the cumulative sum of a stream. Integration is done by taking a sum of the stream with the shifted output.

\begin{verbatim}
integrate : stream -> stream -> cotype.  %mode integrate +A -B.
integrate/def : integrate D B -> add_stream (cons z B) D R -> integrate D R.
\end{verbatim}

We can check that integrating the stream that counts up from $0$ gives the stream $0, 1,3,6,10 ...$.
\begin{verbatim}
main : stream -> cotype.  %mode main -.
main_def : integrate (up z) G -> main G.
\end{verbatim}
\begin{verbatim}
(cons z 
(cons (s z) 
(cons (s (s (s z))) 
(cons (s (s (s (s (s (s z)))))) 
(cons (s (s (s (s (s (s (s (s (s (s z)))))))))) ...)))))
\end{verbatim}

\section{Conclusion}

We have sketched an interpreter for \colfone.  It statically performs mode
checking and uniqueness checking and then executes the program concurrently,
initially with a single process with just one output channel.  During this
execution it also constructs a partial proof object using the constructors
naming the rules.  When new processes are spawned, each is responsible for
sending along a single output channel while receiving from possibly several
input channels.  This matches the computational model of Sax under its
message-passing interpretation \citep{DeYoung20fscd,Pfenning23coordination}.  Our
implementation therefore generates Sax code from the \colfone\ source and then
evaluates it to a certain depth.

The first-order kinds are interpreted as data types, and the clauses 
of first-order kinds are interpreted as data constructors.
The higher-order kinds are interpreted as relations, which 
 are seen as logic programming recipes for 
constructing outputs from inputs with mode specifications.
 Clauses of relations 
are interpreted as concrete steps on how to read inputs and write outputs.
Premises of a clause become concurrently executing subprocesses, and
recursive definitions are seen as a special instance of relations.


Besides a formal description of compilation which is beyond the scope of this
work-in-progress report, our preliminary investigations leave room for multiple
generalizations towards full \colf, such as higher-order terms, dependently
typed terms (not just at the level of proof objects), and the mix of inductive
and coinductive types.  We are encouraged by experiments with our preliminary
compiler and the strong proof-theoretic foundations for both \colf\ and Sax.

\bibliographystyle{eptcs}
\bibliography{citationsconf}

\end{document}